\magnification\magstep1
\parindent=1em
\parskip=0.1ex plus 0.5ex
\hsize=16.5truecm
\vsize=22truecm
\baselineskip=14pt
\font\ttlfnt=cmr10 scaled\magstep 2
\def\d{{\rm d}}
\def\e{{\rm e}}
\def\sign{{\rm sign}}

\def\frac#1#2{{\displaystyle{\displaystyle#1\over\displaystyle#2}}}
\def\p#1#2{\frac{\partial#1}{\partial#2}}
\def\m{M}
\def\p{P}
\def\r{R}
\def\Var{\mathop{\rm Var}}
\def\sx{\sigma(t,x)}
\def\dpar{{\partial}}
\def\mid{\mathop{\vert}}
\def\f{f^{\rm Beta}}
\def\refbgl{29}
\def\refdk{30}
\def\refgrif{31}
\def\refcga{32}
\def\refcgb{33}
\def\refcgc{34}
\def\refbcg{35}
\def\refcox{36}
\def\refkesta{37}
\def\refkestb{38}
\def\refkestc{39}
\def\refkestd{40}
\def\refglau{41}
\def\reflig{42}
\def\refdg{44}
\def\refhg{43}

\input epsf

\noindent{\ttlfnt Large deviations and nontrivial exponents in
coarsening systems}

\bigskip
\noindent I.~Dornic$^{1,2}$ and C.~Godr\`eche$^{1,3}$ 
\bigskip
\noindent
$^{1}$Service de Physique de l'\'Etat Condens\'e,
CEA-Saclay, 91191 Gif-sur-Yvette, France

\noindent
$^{2}$Laboratoire de Physique de la Mati\`ere Condens\'ee, Universit\'e de Nice,
France

\noindent
$^{3}$Laboratoire de Physique Th\'eorique et Mod\'elisation,
Universit\'e de Cergy-Pontoise, France

\bigskip\null\bigskip
\noindent{\bf Abstract.}
We investigate the statistics of the mean magnetisation, of its large deviations and
{\it persistent large deviations} in simple coarsening systems.
We consider more specifically the case of the diffusion equation, of the Ising
chain at zero temperature and of the two dimensional voter model.
For the diffusion equation, at large times, the mean magnetisation has a limit law,
which is studied analytically using the independent interval approximation.
The probability of persistent large deviations, defined as the probability that
the mean magnetisation was, for all previous times, greater than some level $x$, decays algebraically at large times,
with an exponent $\theta(x)$ continuously varying with $x$. 
When $x=1$, $\theta(1)$ is the usual persistence exponent.
Similar behaviour is found for the Glauber-Ising chain at zero temperature.
For the two dimensional Voter model, large deviations of the mean magnetisation are
algebraic, while persistent large deviations seem to behave as the usual persistence
probability.

\vfill

\noindent To be submitted for publication to 
Journal of Physics A

\noindent 24/12/97\medskip
\eject

\noindent{\bf 1 Introduction}
\vskip 12pt plus 2pt

Up to now most studies of persistence in simple nonequilibrium systems have
focussed on the behaviour of the persistence probability at large times and on the
computation of the related persistence exponent [1-28].
The aim of this paper is to broaden the scope of these former
studies by investigating the statistics of more general persistent events.
We will see that consideration of these events leads in particular to the introduction
of new nontrivial exponents.

A simple definition of persistence may be given as follows. 
Let a time-dependent random variable $\sigma(t)$ take only two values
$\pm1$, with some dynamical rule.
Think for instance of $\sigma(t)$ as being the spin at a particular site
in a dynamical Ising model. 
The persistence of this random variable up to time $t$
corresponds to the most extreme situation where it never changed sign. 
In other terms the spin spent all its time in only one of the two possible
phases. 
Note that, by its very definition, this event is non local in time.
The probability of this event, or persistence probability, for most of the systems
mentioned above, decreases algebraically in time, with nontrivial exponents.
The surprise of finding new nontrivial exponents in the dynamics of nonequilibrium
systems motivated to a large extent the interest for the subject.

In this paper we investigate the statistics of {\it large deviations} and of {\it
persistent large deviations} for simple coarsening systems. 
Both are natural generalisations of the concept of persistence.
We apply this study to the case of the diffusion equation, the one dimensional
Glauber-Ising chain at zero temperature, and the two dimensional voter model.

Let us define the mean
`magnetisation' at time
$t$ of the random process
$\sigma(t)$, or `spin' for short, as
$$
\m(t)={1\over t}\int_0^t \d u\,\sigma(u)
,\eqno(1.1)
$$
which is such that $-1\le\m(t)\le1$.
The quantities considered in this work are the following.

\noindent
-- We first define the distribution of the mean magnetisation by
$$
\p(t,x)=P(\m(t)\ge x)
.\eqno(1.2)
$$
This quantity measures the chance for the mean magnetisation to deviate from
its average.\footnote{$^1$}{Hereafter we will only consider
cases with zero average magnetisation, i.e. such that the average
$\langle\sigma(t)\rangle$ of the spin (and therefore of $\m(t)$) over histories is
zero.}
Taking $x>0$ and $t\to\infty$ such that $x$ is 
much larger than the width of the probability density function of $\m$, 
defines the regime of large deviations.
$\p(t,x)$ is then referred to as the probability of large deviations.
 
\noindent
-- We then define
$$ \r(t,x)
=P(\m(u)\ge x,\forall u\le t)
.\eqno(1.3)
$$
This quantity will be hereafter referred to as
the probability of {\it persistent large deviations}.

Persistence, as defined above, corresponds
to the largest deviation such that
$\m(t)=1$ (assuming for instance that $\sigma=1$ initially). 
The persistence probability therefore reads 
$$
\r(t)=P(\sigma(u)=1,\forall u\le t)=\p(t,1)=\r(t,1)
,\eqno(1.4)
$$
thus appearing as a limiting case of the two previous probabilities.

We find, for the models considered in this work, the following results in the long
time regime.

\noindent (i)
For the diffusion equation, $\m(t)$ has a limit law, i.e., $\p(t,x)$ converges to a
limit distribution
$\p_\infty(x)$ when $t\to\infty$. 
Using the independent interval approximation, the moments of this limit
distribution are computed analytically.
Their behaviour at high orders, or the singular behaviour of $\p_\infty(x)$ for
$x\to1$, are related to the persistence exponent $\theta$.
\par \noindent
The probability $\r(t,x)$ is found numerically to behave as
$t^{-\theta(x)}$, with an exponent $\theta(x)$ varying continuously from 0 for $x=-1$,
to $\theta$, the usual persistence exponent, for $x=1$.
(Section 3.) 

\smallskip
\noindent (ii)
For the 1D Ising model at zero temperature,
similar behaviour is found, namely $\p(t,x)\to \p_\infty(x)$
and $\r(t,x)\sim t^{-\theta(x)}$. 
(Section 4.) 

\smallskip
\noindent (iii)
For the 2D voter model, numerical simulations suggest that, 
in the regime of large deviations, $\p(t,x)$ behaves as
$t^{-\tilde\theta(x)}$, with an exponent continuously
varying with $x$, and diverging when $x\to1$. 
They also seem to indicate that $\r(t,x)$ behaves
as $\exp[-J(x)\,(\ln t)^2]$. 
(Section 5.)

We devote the next section to further considerations on large deviations and
persistent large deviations.
A general discussion and generalisations will be given in section 6.

\vskip 14pt plus 2pt
\noindent{\bf 2 Large deviations and persistent large deviations}
\vskip 12pt plus 2pt

\noindent 
Let us comment on the definitions (1.2, 1.3). 

First it is obvious that large deviations reflect a persistence property of the
process. 
Think for instance of an event such that $\m(t)$ takes a value very close to
1, corresponding to a very large deviation.
This is even more true of a {\it persistent} large deviation which is a more
constrained event. 

Let us define the {\it occupation time}
of the phases $(+)$ or $(-)$, i.e. the time spent in the $\sigma=\pm$ phase, by
$$
T_\pm=\int_0^t\d u\,\frac{1\pm\sigma(u)}{2}=t\bigg(\frac{1\pm\m(t)}{2}\bigg)
.\eqno(2.1)
$$ 
In other words, the mean magnetisation $\m(t)$ of a generic spin gives a
measure of the fraction of time that this spin spent in one of the two phases.
Therefore, while a large deviation only requires that a generic spin was {\it most of
the time} in the same phase, a persistent large deviation constrains the spin to
fulfil this condition {\it at all previous times}.
Finally persistence corresponds for the spin to staying {\it always} in the same
phase, hence
$\r(t)=P(T_+=t)$. (Assuming that initially $\sigma=-1$ would lead to the
definition $\r(t)=P(T_-=t)$.) 

Let us now illustrate the previous definitions on the very simple
example of a symmetric random walk on a one-dimensional lattice.
We denote by $\sigma(t)$ the step done by the walker at the discrete time $t$,
where $\sigma=\pm1$ with probability 1/2. 
Starting from the origin at time 0, the position of the walker at time $t$ is given
by $\sum_1^t\sigma(u)$.
Its average position is equal to 0.
The quantity $\m(t)$ introduced above represents the mean speed of the walker. 

The law of the position of the walker is well known.
At large times it is normally distributed around its mean, with a variance proportional
to $t$. 
As a consequence, the density of $\m$ is peaked around $x=0$, with a variance
decreasing as $1/t$. 
The probability of a large deviation, giving a measure of the
chance for the walker to reach a position far away from the origin, is exponentially
small, and is given by (see Appendix A)
\def\eqgd{2.2}
$$ 
\p(t,x)\sim\e^{-t\, I(x)}\qquad (x>0, t\gg1)
,\eqno(\eqgd)
$$
where $I(x)=(1/2)[(1+x)\ln(1+x)+(1-x)\ln(1-x)]$ is an entropy function.
In other words, the law of large numbers holds, the
mean $\m(t)$ converging to its average $\langle \m\rangle=0$, when
$t\to\infty$.
The limit law of $\m$ is a delta peak centered at $x=0$,
and $\p(t,x)\to\p_\infty(x)=H(-x)$, where $H(x)$ is the
Heaviside function.

The persistent large deviation $\m(u)\ge x,\forall u\le t$ corresponds to a situation
where the walker always had a mean speed larger than $x$, i.e., stayed to the right
of the position $x\,t$, between 0 and $t$. 
If $x>0$, $\r(t,x)$
behaves at large times in a similar fashion as in eq.(2.2).
If $x<0$, $\r(t,x)$ has a limit $\r_\infty(x)$ when $t\to\infty$ which is a decreasing
function of
$x$, with a discontinuity at every rational value of $x$ [\refbgl].
In the marginal case $x=0$, it is easy to show that 
$\r(t,x)\approx 1/\sqrt{\pi t}$, for $t$ large.

Finally persistence corresponds for the walker to always stepping in the same
direction. 
The persistence probability is 
$$
\r(t)=\e^{-t\ln 2}
.\eqno(2.3)
$$
Note that $I(1)=\ln 2$.

\smallskip

By analogy with the case of the random walk, we set, for the models studied in the
present work,
\def\eqgdd{2.4}
$$
\p(t,x)\sim\e^{-a(t) I(x)}\qquad (x>0, t\gg1)
,\eqno(\eqgdd)
$$
which defines a function $a(t)$ characteristic of the temporal behaviour of large
deviations, and an entropy function $I(x)$, keeping the same notation as above (see
Appendix A). 
In a similar fashion, setting
$$
\r(t,x)\sim\e^{-b(t) J(x)}\qquad (t\gg1)
,\eqno(2.5)
$$
defines a function $b(t)$ characteristic of the temporal behaviour of the persistent
large deviations.

The time dependence of $a(t)$ and $b(t)$ for the models studied in this work
is summarised in Table 1 and will be discussed in the next sections.

\smallskip 
Let us mention some mathematical references relevant
for this work.
Occupation times have been studied for Markov processes [\refdk], and
for several infinite particle systems [\refgrif, \refcga].
Large deviations for occupation times were studied in [\refcgb, \refcgc, \refbcg,
\refcox]. These references will be commented upon in the course of the paper.
We are not aware of previous references on persistent large deviations.

\vskip 14pt plus 2pt
\noindent{\bf 3 The diffusion equation}
\vskip 12pt plus 2pt

\noindent{\it 3.1 The independent interval approximation and the persistence
exponent}\par 

We first introduce definitions, and remind results, which will be needed in the next
section. 
Consider the equation
$$
\dpar_t\phi({\bf x},t)=\nabla^2\phi({\bf x},t)
,\eqno(3.1)
$$
where $\phi({\bf x},0)$ is gaussian, with zero mean.
Here ${\bf x}$ denotes a point in $d$-dimensional space.
The changes of sign, or zero crossings, of the field $\phi$
at a given space point, occur at times
$t_1$, $t_2,\ldots,t_n$, starting from some time origin, or in the variable
$\tau=\ln t$, at times $\tau_1,\tau_2,\ldots,\tau_n$.

Define, for a given space point ${\bf x}$, the process
$\Phi=\phi/\sqrt{\langle\phi^2\rangle}$.
This process is gaussian and stationary in the time variable $\tau$, i.e. its two
time correlation function
$\langle\Phi(\tau_1)\Phi(\tau_2)\rangle=[\cosh(\tau_2-\tau_1)/2]^{-d/2}$ only depends
on the difference
$\mid\tau_2-\tau_1\mid$ [11, 12]. 
As a consequence, the autocorrelation of the process $\sigma=\sign(\Phi)$ reads [11,
12]
$$
A(\tau)=\langle\sigma(0)\sigma(\tau)\rangle
={2\over \pi}\sin^{-1}\frac{1}{(\cosh \tau/2)^{d/2}}
.\eqno(3.2)
$$
with Laplace transform
$$
\hat A(s)=\int_0^\infty\d \tau\, \e^{-s\tau}A(\tau)
={1\over s}\bigg(1-\frac{d}{2\pi} I_d(s)\bigg)
,\eqno(3.3)
$$
where
$$
I_d(s)=\int_0^\infty\d \tau\, \e^{-s\tau}\frac{\tanh\tau/2}{\sqrt{(\cosh\tau/2)^d-1}}
.\eqno(3.4)
$$

Let us denote by $l_n=\tau_n-\tau_{n-1}$ the intervals between zero
crossings in the $\tau$ variable. 
Considering the intervals as independent
reduces the zero crossing process to a renewal process, entirely described, in the
stationary regime, by $f(l)$, the probability density function of
intervals. 
For such a process the probability $p_n(\tau)$ of having exactly $n$ zero
crossings up to time $\tau$ reads, in Laplace space, 
$$
\eqalign{
\hat p_n(s)&=\frac{1-\hat f}{s\langle l\rangle}\hat f^{n-1}(s),\qquad (n>0)\cr
\hat p_0(s)&=\frac{1}{s}-\frac{1-\hat f}{s^2\langle l\rangle}.
}
\eqno(3.5)
$$
Noticing that $A(\tau)=\sum_{n=0}^{\infty}(-)^np_n(\tau)$, leads to the following
relation between $\hat f$ and $\hat A$ [11, 12]
$$
\hat f(s)=\frac{1-\langle l\rangle s(1-s \hat A)/2}
{1+\langle l\rangle s(1-s \hat A)/2}
=\frac{1-s\sqrt{d/2} I_d(s)}{1+s\sqrt{d/2} I_d(s)}
,\eqno(3.6)
$$
with $\langle l\rangle=\pi \sqrt{8/d}$.

The persistence probability $\r(t)$ is the probability that the field $\phi$ at a given
space point did not change sign up to time $t$.
Equivalently it is the probability that $\sigma(\tau)$ did not flip up to time
$\tau$, i.e. the probability of no zero crossing $p_0(\tau)$. 
At large times it behaves as $t^{-\theta}$, or as $\e^{-\theta \tau}$. 
As a consequence, at large $l$, $f(l)\sim\e^{-\theta l}$, the persistence exponent
appearing as the rightmost pole of 
$\hat f(s)$, $s=-\theta$ [11, 12].

For example, in one dimension:
$$
\eqalign{
I_1(s)
&=\sqrt{2}\int_0^\infty\d \tau\,
\e^{-s\tau}\frac{\cosh\tau/4}{\cosh\tau/2}
=\sqrt{2}\big(\beta(s+1/4)+\beta(s+3/4)\big)\cr
&=\sqrt{2}\sum_{p=0}^\infty(-)^p\bigg(\frac{1}{s+1/4+p}+\frac{1}{s+3/4+p}\bigg)
,}
\eqno(3.7)
$$
with $I_1(0)=2\pi$.
The function $\beta(x)$ is related to $\psi(x)$, the logarithmic
derivative of the gamma function, by
$$
\beta(x)={1\over 2}\big[\psi\big({x+1\over 2}\big)+
\psi\big({x\over 2}\big)\big]
.\eqno(3.8)
$$
\noindent In two dimensions:
$$
\eqalign{
I_2(s)
&=\int_0^\infty\d \tau\,
\frac{\e^{-s\tau}}{\cosh\tau/2}
=2\beta(s+1/2)\cr
&=2\sum_{p=0}^\infty(-)^p\frac{1}{s+1/2+p}
,}
\eqno(3.9)
$$
with $I_2(0)=\pi$.
The largest zero of ${1+s\sqrt{d/2} I_d(s)}$ is found to be at $s=-\theta$, with
$\theta=.120 327 978 84\ldots$, for $d=1$ and $\theta=.1862 210 712 97\ldots$, for
$d=2$.

\medskip\noindent
{\it 3.2 Statistics of the mean magnetisation\ }\par
Our concern, in this section, is the determination of the distribution of the random
variable
$\m(t)$, at large times. 
We denote by $t$, or by $\tau$ in the logarithmic scale, the observation time and by
$\lambda$ the `backward recurrence time', i.e. the length of time measured backwards
from $\tau$ to the last crossing event before $\tau$: $\lambda=\tau-\tau_{n}$.
The probability distribution of $\lambda$  in Laplace space reads, in the stationary
regime,\footnote{$^2$}{Let us note that the
age of the system considered in [21] is just equal to $t-t_n$. It is related
to the scaling variable $t_n/t=\e^{-\lambda}$, the distribution of which is known
in the case considered here.}
$$
\hat q(s)=\frac{1-\hat f(s)}{s\langle l\rangle}
.\eqno(3.10)
$$
We have, assuming that $\sigma(t)=1$,
$$
\m(t)={1\over t}\int_{0}^t\d u\, \sigma(u)=
{1\over t}\big(t-t_n-(t_n-t_{n-1})+\cdots\big)
=1-2\xi
,\eqno(3.11)
$$
where
$$
\xi={t_n\over t}-{t_{n-1}\over t}+\cdots
=\e^{-\lambda}\big(1-\e^{-l_n}+\e^{-l_n-l_{n-1}}-\cdots\big)
=\e^{-\lambda} X_n
.\eqno(3.12)
$$
Assuming that $\sigma(t)=-1$ leads to $\m(t)=2\xi-1$.
Note that $\xi=T_{\mp}/t$, according to the sign of $\sigma(t)$, i.e. $\xi$ is
the fraction of time spent in the `wrong' phase (cf eq. (2.1)).
The random variable
$X_n=1-\e^{-l_n}+\e^{-l_n-l_{n-1}}-\cdots$ obeys the recursion relation
$X_n=1-\e^{-l_n}X_{n-1}$.
It is therefore recognised as a Kesten variable [\refkesta, \refkestb, \refkestc,
\refkestd].

To summarize, in the limit $t\to\infty$, the three equations
$$
\eqalignno{
\m&=\pm(1-2\xi),&(3.13\, {\rm a})\cr
\xi&=\e^{-\lambda} X,&(3.13\,{\rm b})\cr
X&=1-\e^{-l}X, &(3.13\,{\rm c})
}
$$
contain the relevant information for the determination of 
$f_\m(x)=-\d \p_\infty(x)/\d x$, the distribution of $\m$.
Equation (3.13 c) should be understood as an equality in distribution.

The determination of the probability density of the Kesten variable $X$
(and hence of $f_\m(x)$) for any given distribution of intervals $f(l)$ is known in
general as a hard problem, and does not seem feasible in the present case.
However, from this set of equations we are able to extract the following information.

\noindent 
(i) We perform a local analysis of $f_\m(x)$, for $x\to 1$.
\par\noindent 
(ii) We compute the moments $\langle \m^k\rangle$ of $f_\m(x)$.
\par\noindent 
(iii) We solve the set of equations (3.13 a, b, c) for the case of an exponential 
distribution of intervals, $f(l)=\theta\e^{-\theta l}$, proportional to the tail
of the true distribution given, in Laplace space, by eq. (3.6).

Let us analyse the local behaviour of $f_\m(x)$ in the persistence region $x\to1$ (a
similar analysis would hold in the limit $x\to-1$). Then $\xi\to0$, i.e,
$\lambda\to\infty$, with
$X$ finite. These conditions define the persistence region, where $s\approx-\theta$.
Therefore $f(l)\approx a\e^{-\theta\,l}$ and 
$q(\lambda)\approx a\e^{-\theta\,l}/\langle l\rangle\theta$, where $a$ is the
residue of $\hat f(s)$ for $s=-\theta$. 
Consider the Mellin transform of the law of $\xi$, $\langle\xi^s\rangle$.
From (3.13b) one gets
\def\eqMellin{3.14}
$$
\langle\xi^s\rangle=\hat q(s)\langle X^s\rangle
,\eqno(\eqMellin)
$$
where $\hat q(s)$ is the Laplace transform (3.10).
In this regime, one has 
$\langle\xi^s\rangle\approx b/(s+\theta)$, with $b$ given by
$$
b=\frac{a}{\langle l\rangle\theta}\langle X^{-\theta}\rangle
.\eqno(3.15)
$$
By inversion of the Mellin transform eq. (\eqMellin) one obtains the behaviour of the
distribution of
$\xi$, hence that of $\m$ in the persistence region $x\to1$.
One finds
\def\eqlocal{3.16}
$$
f_\m(x)\approx 2^{-\theta-1}b (1-x)^{\theta-1}\qquad (x\approx 1)
.\eqno(\eqlocal)
$$
As a consequence, for large $k$ one has 
\def\eqmk{3.17}
$$
\langle \m^k\rangle\approx 2^{-\theta} b\, \Gamma(\theta)\, k^{-\theta}
.\eqno(\eqmk)
$$  

Note that the determination of $b$ requires that of $\langle
X^{-\theta}\rangle$, which is unknown.
A numerical estimate of the amplitude $b$ can nevertheless be given, as follows.
Using the method given in Appendix B, we computed the numerical values of the
first 50 moments of $\m$ in one and two dimensions, in the independent interval
approximation.  
By extrapolating these results we find $2^{-\theta} b\, \Gamma(\theta)\approx.870$ for
$d=1$, and $2^{-\theta} b\, \Gamma(\theta)\approx.8424$ for $d=2$. 
In figure~1 a plot of $k^{-\theta}\langle \m^k\rangle$ versus $1/k$, for $d=2$, is given,
showing the approach to the limit amplitude.

\midinsert
\epsfysize=70truemm
$$\epsfbox{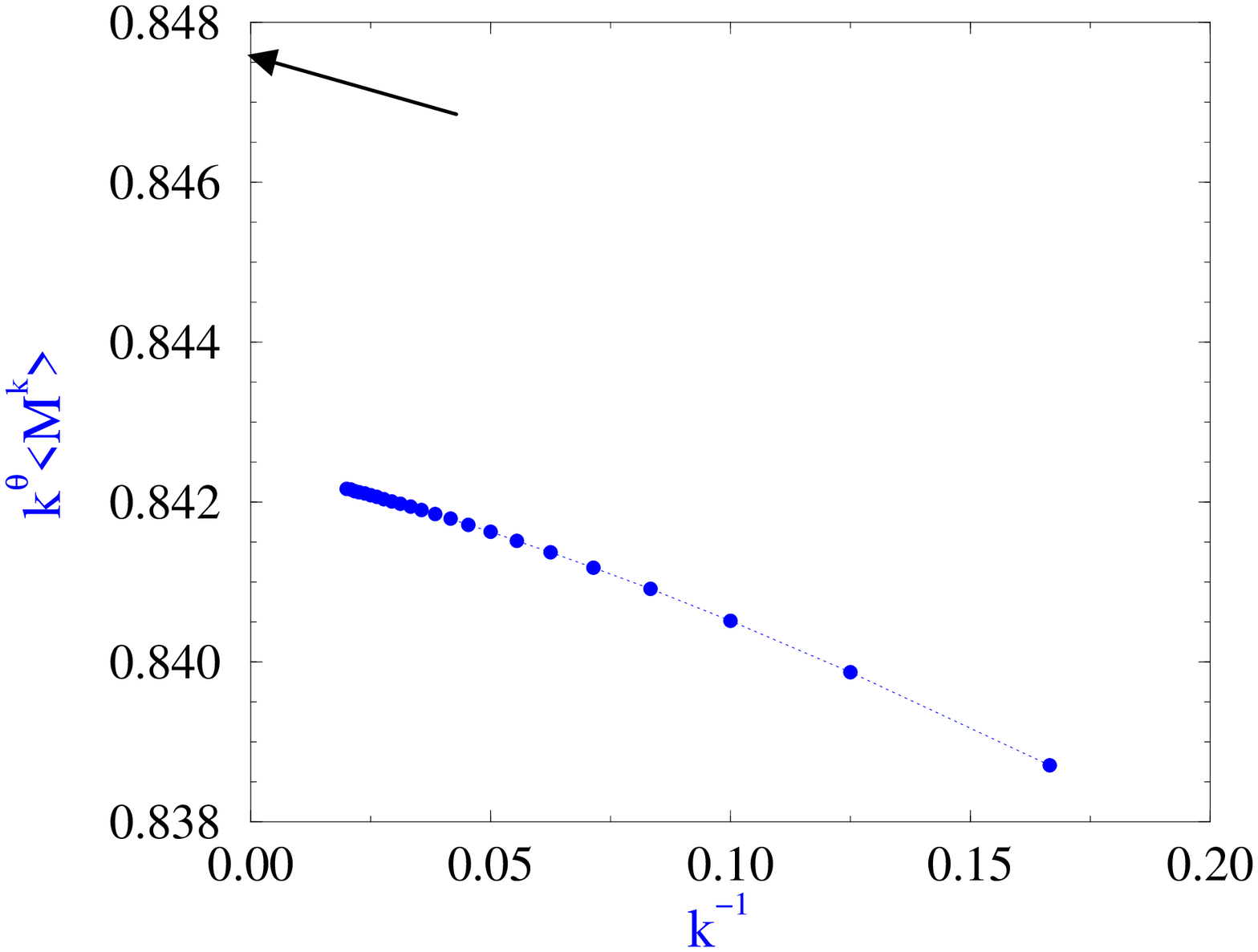}$$
{
{\bf Figure 1:} Plot of $k^{\theta}\langle M^k\rangle$ ($k=2,4,\ldots,50$)
for the 2D diffusion equation in the independent interval approximation, 
versus $k^{-1}$, showing the approach to the limit amplitude .8424.
Arrow: value of the limit amplitude .8476 for the beta law (3.18).
}
\endinsert

We are naturally led to compare the distribution $f_\m(x)$ to a beta law on
$(-1,1)$, with the same singular behaviour in the region $x\approx1$, 
\def\eqsing{3.18}
$$
\f(x)=B^{-1}(1/2,\theta)\,(1-x^2)^{\theta-1}
,
\eqno(\eqsing)
$$
with 
$$
B(1/2,\theta)=\frac{\Gamma(1/2)\Gamma(\theta)}{\Gamma(1/2+\theta)}
.\eqno(3.19)
$$
The even moments of this law are given by
$$
\mu_{k}=\frac{B((k+1)/2,\theta)}{B(1/2,\theta)}
.\eqno(3.20)
$$
The odd moments are zero, by construction.
At large orders,
\def\eqmuk{3.21}
$$
\mu_k\approx 2^\theta\frac{\Gamma(1/2+\theta)}{\Gamma(1/2)}\,k^{-\theta}
\qquad (k\gg 1) 
.\eqno(\eqmuk)
$$
Defining the ratio of local amplitudes $A$ by $f_\m\approx A \f$, for
$x\approx1$, yields
\def\eqA{3.22}
$$
A=\lim_{k\to\infty}\frac{\langle \m^k\rangle}{\mu_k}
=b\,B(\theta,\theta+1)
,
\eqno(\eqA)
$$
using (\eqmk, \eqmuk).
Hence using the numerical estimates of $b$ given above,
we find $A\approx.982$ for $d=1$, and $A\approx.9938$
for $d=2$. 
Table 2 gives the values of the first moments $\langle \m^k\rangle$, compared to the
moments of the beta law (\eqsing).

The resemblance of $f_\m(x)$ to the beta law (\eqsing) is enhanced by the fact that 
the solution of (3.13 a, b, c) for an exponential distribution of intervals
$f(l)=\theta\e^{-\theta l}$, proportional to the tail of the true distribution $f(l)$
given, in Laplace space, by eq. (3.6), is precisely given by (\eqsing) (see Appendix B).
This demonstrates the dominance of the tail of $f(l)$ for the determination of $f_\m(x)$.

We also computed the probability distribution of the mean magnetisation obtained by
numerical integration of eq. (3.1), for $d=1$.
This distribution is also found to be very close to $\f$.

\smallskip
In summary, the mean magnetisation $\m(t)$ has a limit
distribution $\p_\infty(x)$ when $t\to\infty$. 
In other words there is no law of large numbers for the random process
$\sigma(t)$, and absence of ergodicity. 
This distribution is found to be extremely
close to the beta distribution (\eqsing).
As long as $\theta<1$, the density $f_\m(x)$ diverges for $x\to\pm1$.
Therefore the most probable values of $\m$ are near $-1$ and $1$, while the average
$\langle \m\rangle=0$. 
The function $I(x)$ diverges when $x\to1$.
This signals the crossover to the persistence regime.
The function $a(t)$ introduced in (2.4) is formally of order
unity. 
By extension, we will still speak of large deviations when $t\to\infty$ and $x\approx1$.

\medskip\noindent
{\it 3.3 Persistent large deviations of the mean magnetisation\ }\par

We performed numerical simulations of the diffusion equation eq.(3.1) in 1D,
for a system size equal to $10^6$, starting from a random initial condition.
At large times one observes an algebraic decay of the probability of persistent large
deviations $\r(t,x)$ of the form
$$
\r(t,x)\sim t^{-\theta(x)}\qquad (-1\le x\le1)
,\eqno(3.23)
$$
which corresponds to the behaviour $b(t)\sim \ln t$ for the function
defined in (2.5).
The exponent $\theta(x)$ is to be identified to $J(x)$ defined in eq. (2.5).

Figure 2 gives a plot of the probability of persistent large deviations $\r(t,x)$
for $x=-.8$. 
The usual persistence probability $\r(t)$ is also plotted, for comparison.
The third curve corresponds to $\r(t,x,y)$ defined at the end of section 6 (see the
comment there).
\par

\midinsert
\epsfysize=70truemm
$$\epsfbox{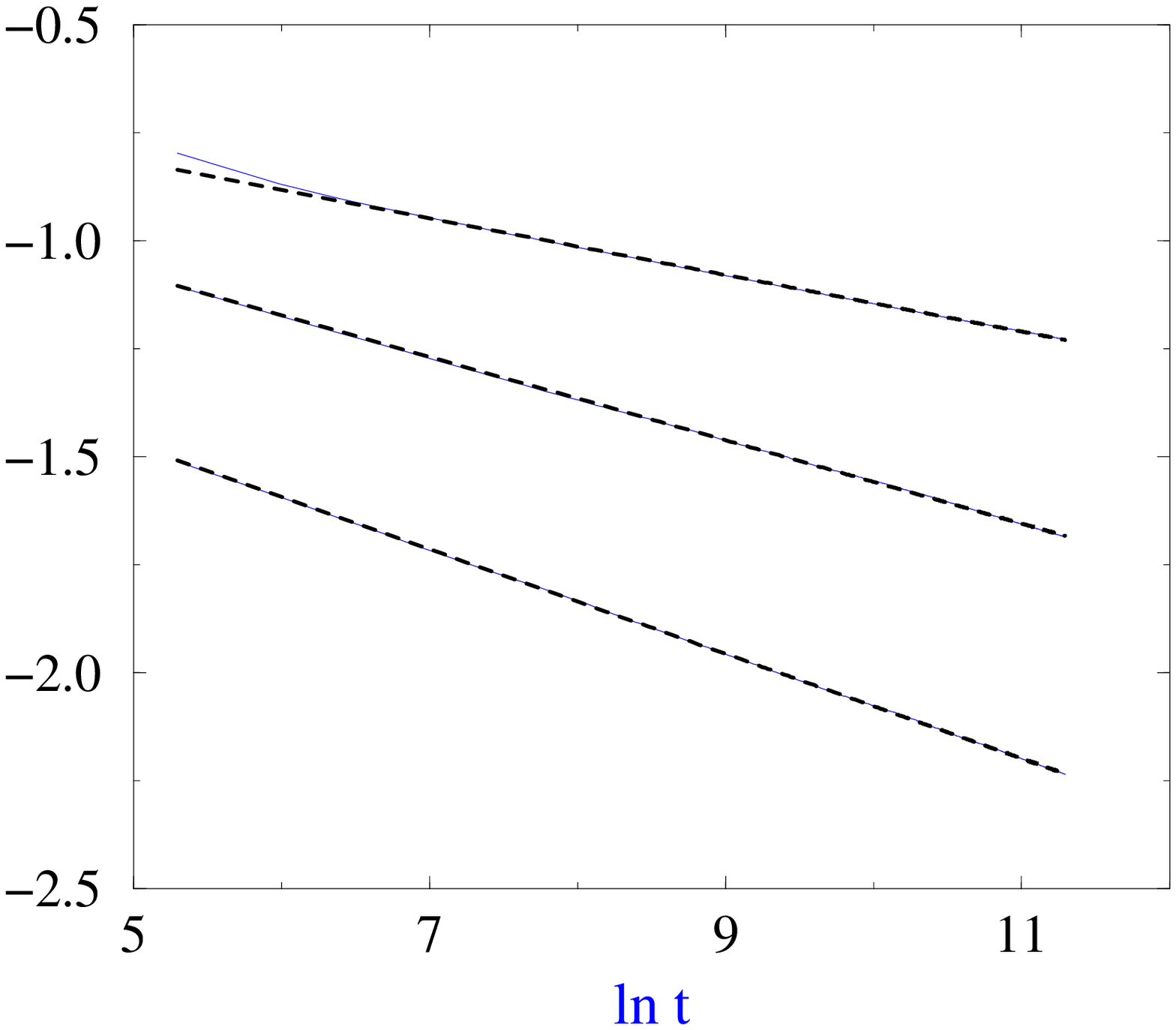}$$
{{\bf Figure 2:} Persistence probability $R(t)$, probability of persistent large deviations
$R(t,x)$, and $R(t,x,y)$ (see section 6), for the 1D diffusion equation. 
(System size=$10^6$.)
From bottom to top: $R(t)$ (slope$=-.121$), $R(t,-.8)$ (slope$=-.096$), $R(t,-.8,-.8)$
(slope$=-.065$). In dashed: regression lines.
}
\endinsert

Figure 3 shows a plot of the exponent $\theta(x)$ for $-1\le x\le1$.
The exponent varies continuously from 0, for $x=-1$, to the value of the usual
persistence exponent $\theta\approx.121$, for $x=1$.
(We recall that the value of $\theta\approx.121$ obtained by numerical 
integration of (3.1) is
slightly larger than the value of the exponent obtained by the independent 
interval approximation [11, 12].)
We will further comment on these results in section 6.

\midinsert
\epsfysize=70truemm
$$\epsfbox{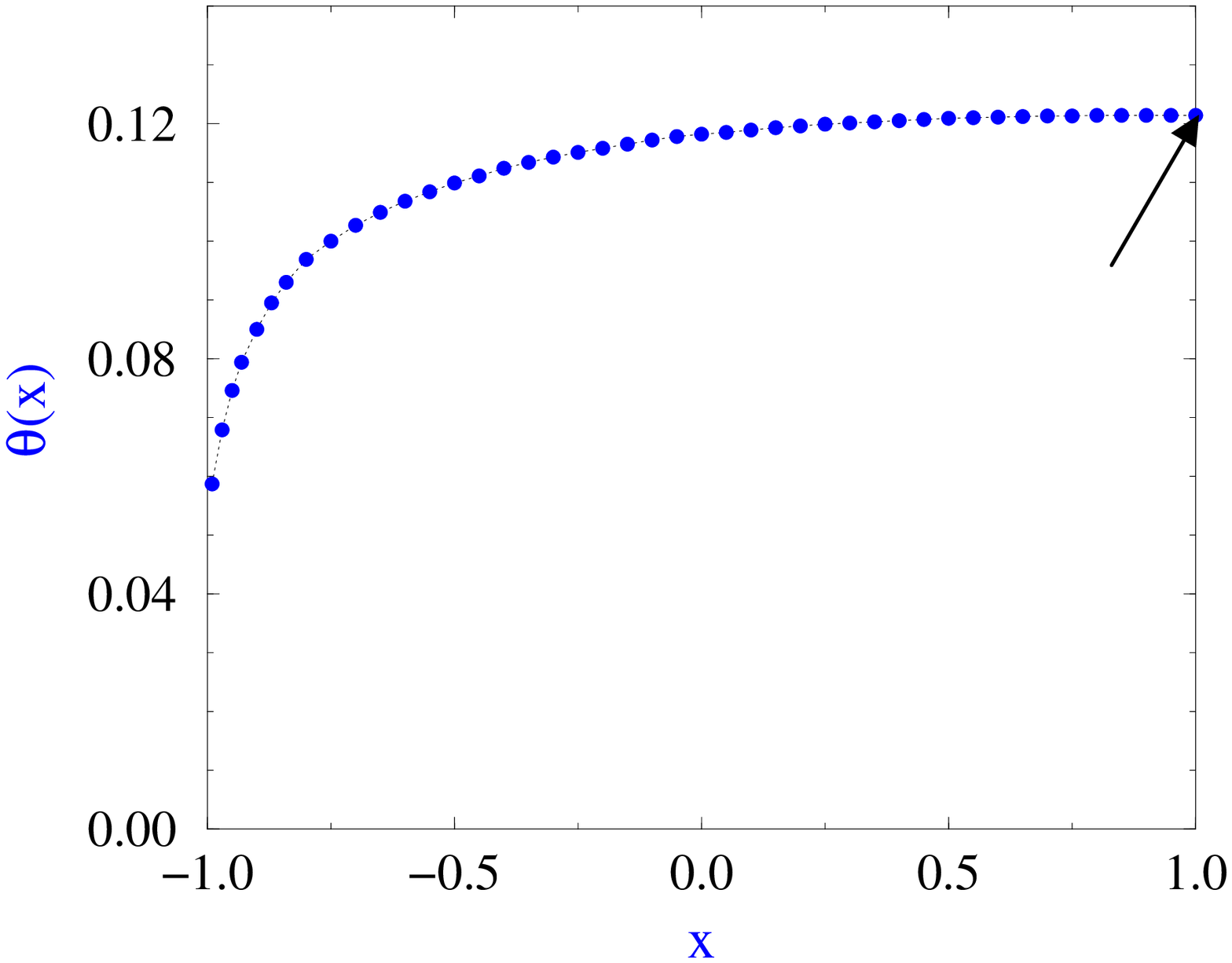}$$
{{\bf Figure 3:} Exponent $\theta(x)$ for the 1D diffusion equation. 
Arrow: the usual persistence exponent $\theta\approx.121$. }
\endinsert

Let us finally mention that similar results as those presented in figures 2, 3
are found in two dimensions.
\vskip 14pt plus 2pt
\noindent{\bf 4 The Glauber-Ising chain}
\vskip 12pt plus 2pt

We studied the Ising chain at zero temperature with the following dynamics
[\refglau]. On each site of the 1D lattice, values of the
spin $\sigma=\pm$ are initially distributed randomly.
At each time step a site is picked at random.
The spin on this site takes the value of one of its neighbours, chosen at random.

We performed numerical simulations on a system of size $L=10^6$.
As for the case of the diffusion equation, $\p(t,x)$ has a limit distribution when
$t\to\infty$.
This distribution is very close to a beta law corresponding to the persistence
exponent $\theta=3/8$. 
The analytical study of $\p(t,x)$ will be given elsewhere.
In particular it is easy to understand why this distribution converges to a limit
when $t\to\infty$.
For instance $\langle M^2\rangle=\hat A(1)$, where $\hat A(s)$ is the Laplace
transform of the autocorrelation function
$A(\tau)=\langle\sigma(0)\sigma(\tau)\rangle$ with respect to the logarithmic time
$\tau=\ln t$ (see eq. (B.7))  [\refdg].

The probability of persistent large deviations
$\r(t,x)$ decays algebraically, with an exponent continuously varying with $x$
(figures 4 and 5). 
For $x=1$ the usual persistence exponent $\theta=3/8$ is recovered.

\midinsert
\epsfysize=70truemm
$$\epsfbox{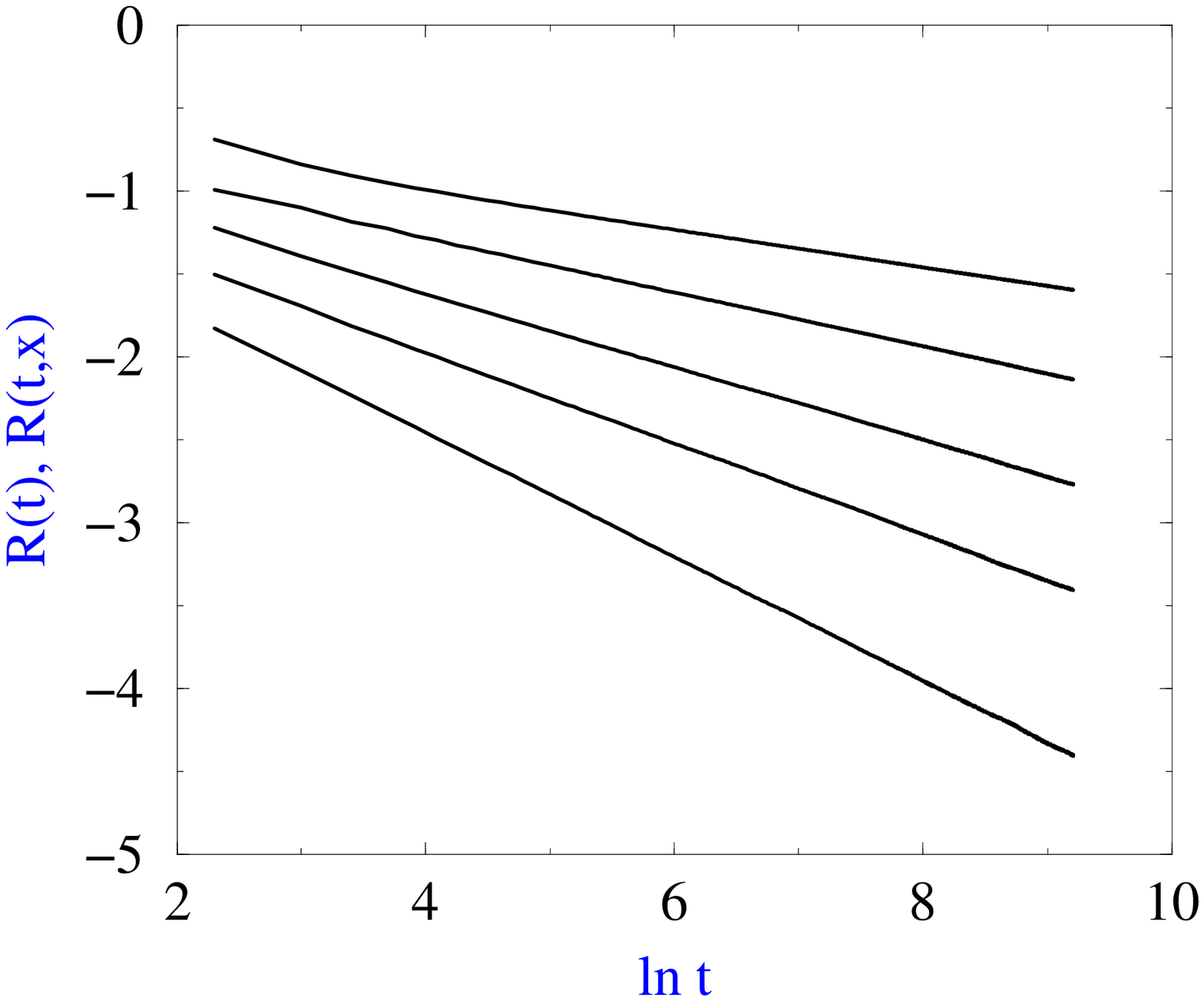}$$
{{\bf Figure 4:} Probability of persistent large deviations $R(t,x)$ for the
1D Ising model.  
(System size=$10^6$.)
From bottom to top: $R(t)$, $R(t,.5)$, $R(t,0)$, $R(t,-.5)$, $R(t,-.8)$. 
}
\endinsert

\midinsert
\epsfysize=70truemm
$$\epsfbox{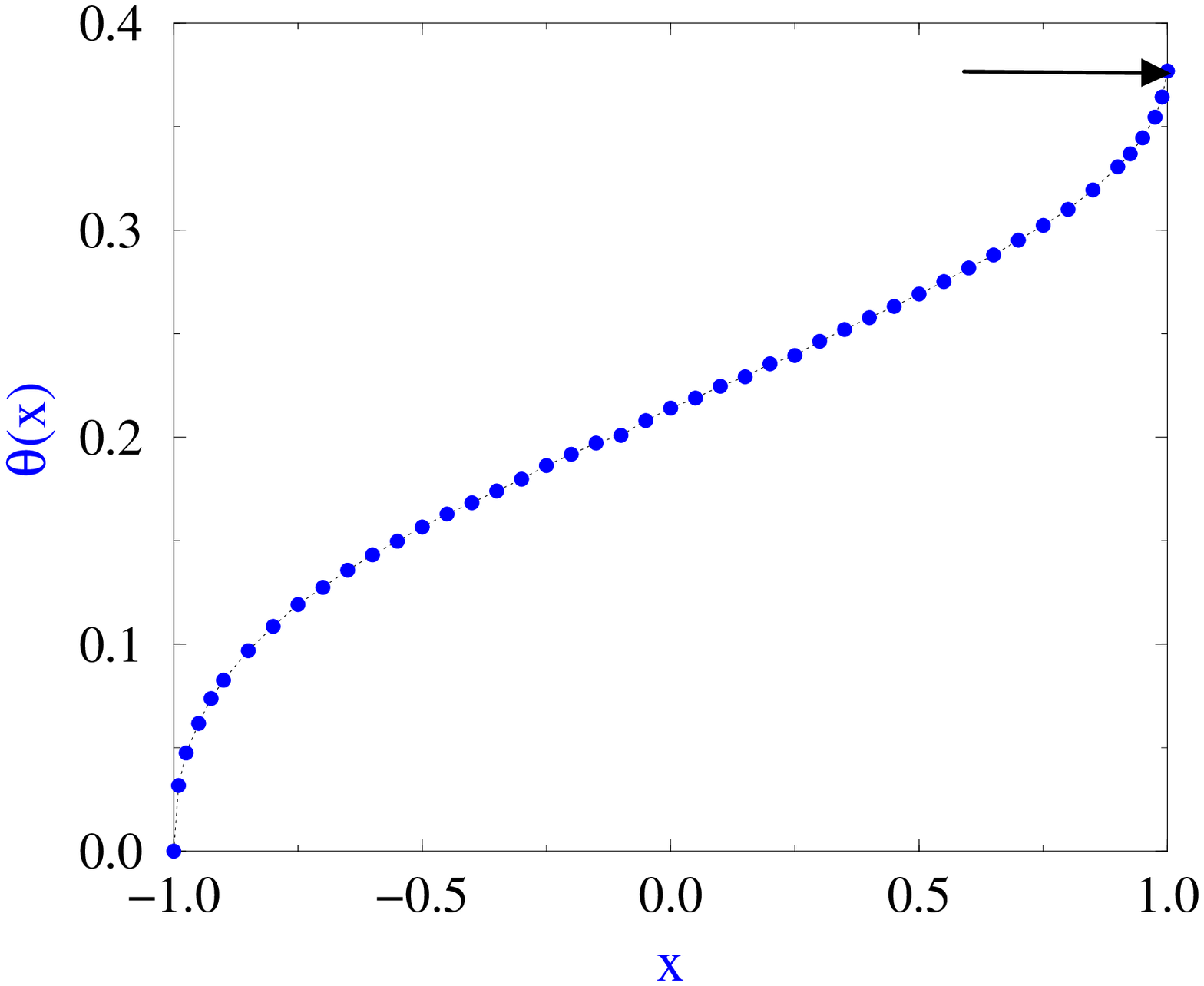}$$
{{\bf Figure 5:} Exponent $\theta(x)$ for the 1D Ising model. 
Arrow: the usual persistence exponent $\theta=3/8$. }
\endinsert

\vskip 14pt plus 2pt
\noindent{\bf 5 The 2D voter model}
\vskip 12pt plus 2pt

The voter model  is defined as follows [\reflig].
On each site of a $d-$dimensional lattice, opinions of
a voter or values of a spin
$\sigma=1, 2,\ldots,q$ are initially distributed randomly.
At each time step a site is picked at random.
The voter on this site takes one of the opinions of its $2D$ neighbours, with
equal probabilities.
Hence the rules of the dynamics of the voter model are a simple
generalisation of those of the 1D Ising model at zero temperature. 
In particular the 1D voter model is identical to the Glauber-Ising chain.

Earlier references to $\p(t,x)$ or to the occupation time $T_+$ for the
voter model may be found in [\refcga, \refcgc, \refbcg, \refcox]. 
The function $a(t)$ appearing in the large
deviation expression eq. (2.4) is  related to the variance of the occupation time
$T_+$ [\refcga, \refcgc, \refbcg], hence to the two time correlation function of the
process [\refdg], by (see Appendix A)
$$
a(t)\sim \frac{t^2}{\Var T_+}\sim{1\over \Var M}
.\eqno(5.1)
$$  
In one dimension, $\Var M=$ const., as mentioned in section 4,
hence $a(t)={\cal O}(1)$. 
The convergence in distribution of $T_+/t$ in one dimension was shown in [\refcga].
In two dimensions, $\Var M\sim 1/\ln t$ [\refcga, \refcgc, \refbcg, \refcox, \refdg], hence $a(t)\sim \ln t$.
Therefore the rate at which the distribution of $\m$ gets peaked is very slow.
It was conjectured in [\refcgc, \refbcg] that this estimate is exact,
i.e. that $a(t)=\ln t$, hence that, in $d=2$, $\p(t,x)$ should have algebraic decay
when $t\to\infty$.
For $d>2$, $a(t)$ is respectively equal to $\sqrt{t}$, $t/\ln t$, $t$
for $d=3,\, 4$ and $d >4$ [\refcga, \refcgc, \refbcg, \refcox, \refdg].

We performed numerical simulations of the 2D voter model, for system sizes up to $(4000)^2$, with $q=2$.
These simulations suggest that $\p(t,x)$ behaves as
$t^{-\tilde\theta(x)}$, with an exponent continuously
varying with $x$ ($x>0$), and to be identified to $I(x)$ defined in eq. (2.4).
We will present the numerical analysis of the scaling in [\refdg].

The numerical results also seem to indicate that $\r(t,x)$ behaves
as $\exp[-J(x)\,(\ln t)^2]$,
reminiscent of the behaviour of the usual persistence probability [16, \refhg].
Hence the function  $b(t)$ introduced in (2.5) is equal to $(\ln t)^2$.
It is nevertheless rather hard to conclude, on the basis of our numerical simulations.

We conjecture that $b(t)$ is proportional to $N(t)$, the average number of particles 
in the dual particle system (diffusing coagulation, or equivalently reaction diffusion A+A$\to$A, with a local source). 
$N(t)$ is equal respectively to $\ln t$, $(\ln t)^2$, 
$\sqrt{t}$, $t/\ln t$, $t$
for $d=1,\,2,\,3,\, 4$ and $d >4$ [\refcga, \refcgc, \refbcg, \refcox, \refhg]. 
These results have a clear intuitive interpretation.
When the dimension of space increases, particles interact
less strongly since they have more and more space to explore before meeting. 
As a consequence, they are less correlated and their average number
increases.
In high enough dimension, the reaction between particles becomes irrelevant,
hence $N(t)$ becomes proportional to $t$,
reflecting the total independence of
the particles.
Note that above two dimensions, $a(t)$ is equal to $N(t)$ (and therefore to $b(t)$).

The voter model therefore interpolates between the case of the Glauber-Ising chain seen in
previous section, if $d=1$, and the case of the random walk of section 2,
if $d>4$.

\vskip 14pt plus 2pt
\noindent{\bf 6 Discussion and conclusion}
\vskip 12pt plus 2pt

The most striking conclusions that may be drawn from this study are, from our point
of view,

\noindent (i) the existence of a limit distribution for $\p(t,x)$, the probability
distribution of the mean magnetisation, for the diffusion equation in all dimensions
and the 1D Ising chain,

\noindent (ii) the appearance of families of exponents in the temporal decay
of $\r(t,x)$, the probability of persistent large deviations, for the diffusion
equation and the 1D Ising chain.\break
Finally there are indications that the large deviations $\p(t,x)$ for the 2D voter
model are algebraic.  
This extends the scope of former studies on the persistence exponents found in
coarsening systems.

A number of comments are in order.

The time dependence of $a(t)$ and $b(t)$, defined in (2.4, 2.5), is summarised
in Table 1. One observes that for the random walk, $a(t)$ and $b(t)$ are
proportional to $t$, while these functions are slowly increasing with time for the
coarsening systems considered in this work, or even constant.
This may be interpreted as follows.
In the case of the random walk, $\m(t)$ is given by a sum of $t$ independent
random variables. 
In the coarsening systems studied here, values of the spin at a fixed position at
different times are strongly dependent random variables.
Thus $a(t)$ and $b(t)$ measure in some sense the effective number of independent
variables in the system. 
This is clear for $a(t)$, from its very definition given in Appendix A, and for
$b(t)$, at least for the voter model, from the discussion given at the end of section~5.

Thus the mechanism by which $\r(t,x)$ --and as a consequence $\r(t)$, the usual
persistence probability-- have algebraic decay at large times, for the diffusion
equation and the 1D Ising model, can be traced back to the logarithmic behaviour of
$b(t)$ defined in eq. (2.5). 
The same comment holds for $\p(t,x)$ and $a(t)$ for the 2D Voter
model.

Unfortunately the exact computation of the exponents seems a difficult
task, since this amounts to computing the `entropy' functions $I(x)$ or $J(x)$.
Already computing the usual persistence exponent, corresponding in the present
framework to taking the limit $x\to1$, is in general difficult.
Moreover, even for the simple random walk, the probability $\r(t,x)$ is a nontrivial
mathematical object.
At least for the diffusion equation, for which it is possible to get analytic results
for $\p(t,x)$ in the independent interval approximation (see section 3), one could hope
of computing $\r(t,x)$. 
Let us note that some aspects of this work, for instance the
interpretation of the exponents as entropy functions, are reminiscent of the
multifractal formalism.

One also observes that for coarsening systems, when $x\to 1$, $I(x)$ diverges, while
$J(x)$ converges to a constant.
The divergence of $I(x)$ signals the crossover from large deviations ($\p(t,x)$) to
persistence ($\r(t)$).
The convergence of $J(x)$ shows that $\r(t,x)$ is a natural generalisation of the
persistence probability $\r(t)$, $b(t)$ encoding the type of decay of the persistence
probability, being algebraic or not.

We can enhance the difference between $\r(t,x)$ and $\p(t,x)$ as follows.
First define the new random variable
$$
\sx=\sign(\m(t)-x)
,\eqno(5.1)
$$
which is an indicator of whether the mean magnetisation $\m(t)$ at time $t$ is above
or below the level $x$.
One has
$$
\p(t,x)=\big\langle\frac{1+\sx}{2}\big\rangle
.\eqno(5.2)
$$
Then $\r(t,x)=P(\m(u)\ge x,\forall u\le t)$ is just the persistence probability of
this random variable.
Therefore
$$
\r(t,x)
=P(\sigma(u,x)=1, \forall u\le t)
=\big\langle\frac{1+\sigma(t_1,x)}{2}
\frac{1+\sigma(t_2,x)}{2}\cdots\frac{1+\sigma(t_n,x)}{2}\big\rangle
,\eqno(5.3)
$$
(taking a discrete set of intermediate times, then letting $n\to\infty$) which shows
that $\r(t,x)$ is a highly non local function of time.

One may generalise the present approach by progressively `thinning' large deviations,
and tracking rarer and rarer events. 
Let us define
$$
\m(t,x)={1\over t}\int_0^t \d u\sigma(u,x)
,\eqno(5.4)
$$
($-1\le \m(t,x)\le1$) and the corresponding probabilities
$\p(t,x,y)=P(\m(t,x)\ge y)$
and
$\r(t,x,y)=P(\m(u,x)\ge y,\forall u\le t)$.

First consider the random walk.
Take $x=0$ for simplicity.
Then $\m(t,0)$ is simply related to the fraction of time the walker spends on the
right side of the origin.
The limit distribution of this quantity (when $t\to\infty$) is given by the arc
sine law [45]. 

We computed $\r(t,x,y)$ on the diffusion equation, and on the Ising chain. 
For example, figure 2 shows $R(t,x,y)$, with $x=y=-.8$, for the 1D diffusion
equation. 
Again algebraic decay is observed.

Let us point out that this progressive thinning of large deviations implies
probing the system by events which are always more non local in time.
This questions the possibility of the existence of an infinite number of exponents in
temporal quantities measured on strongly interacting systems.

\bigskip\noindent
{\bf Acknowledgements\ }
\medskip 

We thank J.P. Bouchaud, H. Chat\'e, S. Majumdar and J. Neveu for interesting conversations.
We are deeply indebted to J.M. Luck for pointing out to us the role played by the
Kesten variable in section 3.

\vfill\eject

\noindent{\bf Appendix A}
\vskip 12pt plus 2pt

This appendix provides an explanation of the large deviation expressions
(\eqgd) and (\eqgdd).

\medskip\noindent
{\it Independent random variables}
\medskip

Consider $Y=t\m=\sum_1^t\sigma(u)$, where the $\sigma$ are independent identically
distributed random variables. 
The generating function of moments of $Y$ is
$$
\langle \e^{sY}\rangle=\hat f_Y(s)=\e^{tK(s)}
,\eqno({\rm A.}1)
$$
where $\hat f_Y(s)$ is the Laplace transform of the density of $Y$,
and $K(s)$ is the generating function of cumulants of the random variable $\sigma$.
Inverting the Laplace transform yields
$$
f_Y(y)=\int \frac{\d s}{2\pi i}\e^{-s\,y+t\,K(s)}
=\int \frac{\d s}{2\pi i}\e^{-t[s\, x-K(s)]}
,\eqno({\rm A.}2)
$$
where $x=y/t$.
For $t\to\infty$ we use the saddle point method to evaluate the integral.
At the saddle point, $K'(s_c)=x$. 
Defining
$$
I(x)=s_c\,x-K(s_c)
,\eqno({\rm A.}3)
$$
yields $f_Y(y)\sim\e^{-t\,I(x)}$.
Finally
$$
P(t,x)\sim \e^{-t\,I(x)}
,\eqno({\rm A.}4)
$$
which is (\eqgd).
Note that $I(x)$ is the Legendre transform of $K(s)$.

Let us apply this general formalism to the case of the random walker (see
section 2).
Then $\langle \e^{s\sigma}\rangle=\cosh s$,
and $K(s)=\ln \cosh s$.
At the saddle point $x=K'(s_c)=\tanh s_c$, hence
$$
s_c={1\over 2}\ln\frac{1+x}{1-x}
.\eqno({\rm A.}5)
$$
Noting that $\cosh s_c=(1-x^2)^{-1/2}$, leads to 
$I(x)=(1/2)[(1+x)\ln(1+x)+(1-x)\ln(1-x)]$.

\medskip
\noindent
{\it Correlated random variables}

Consider the generating function of the cumulants of $\m$, denoted by $c_n$,
$$
K_\m(s)=\ln \langle \e^{s\m}\rangle
=\sum\frac{c_n}{n!} s^n
.\eqno({\rm A.}6)
$$
Assume that, when $t\to\infty$, the $c_n$ scale as [\refcgc]:
$$
c_n\approx\frac{b_n}{[a(t)]^{n-1}}\quad (t\to\infty)
,\eqno({\rm A.}7)
$$
where $a(t)$ diverges with $t$, and the $b_n$ are constants.
Hence $a(t)\sim1/\Var M$.

We now consider $Y=a(t)\m$.
Then
$$
\frac{1}{a(t)}\ln \langle\e^{sY}\rangle=
\frac{1}{a(t)}K_\m(s a(t))
\approx\sum\frac{b_n}{n!}s^n\equiv K(s)\quad (t\to\infty)
.\eqno({\rm A.}8)
$$
Hence at large times
$$
\langle \e^{sY}\rangle\approx
\e^{a(t) K(s)}
,\eqno({\rm A.}9)
$$
from which one gets
$$
f_Y(y)=\int \frac{\d s}{2\pi i}\e^{-s\,y+a(t)\,K(s)}
=\int \frac{\d s}{2\pi i}\e^{-a(t)[s\, x-K(s)]}
,\eqno({\rm A.}10)
$$
where $x=y/a(t)$.
Continuing as above, we obtain, at large times
$$
\p(t,x)\sim\e^{-a(t) I(x)}
,\eqno({\rm A.}11)
$$
which is (\eqgdd).
Again, $I(x)$ is given by (A.3) and $K'(s_c)=x$, 

Here, the role of $a(t)$ parallels that played by $t$
for the former case of independent variables.
This function can be therefore interpreted as the effective number of independent variables, in the case where the spins $\sigma$ at different times are correlated.

Note that in the cases of the diffusion equation or the 1D Ising-Glauber chain, 
$a(t)={\cal O}(1)$, because all cumulants become constant when $t\to\infty$.

\vskip 12pt plus 2pt
\noindent{\bf Appendix B}
\vskip 12pt plus 2pt

In this Appendix we first show how to compute the moments of the distribution
of the mean magnetisation for the diffusion equation.
We then solve the set of equations (3.13 a, b, c) for the case of an exponential 
distribution of intervals, $f(l)=\theta\e^{-\theta l}$, proportional to the tail of
the true distribution given by eq. (3.6). 

Using eqs. (3.13 a-c), the moments $\langle \m^k\rangle$ can be computed recursively
as follows. The computation is done in three steps.

\noindent (i)
From (3.13 c) one computes the moments of $X$ from those of $\e^{-l}$,
i.e. as functions of the coefficients $\hat f_k$, recursively, where $\hat f_k$
denotes $\hat f(s)$ for integer values of the argument.
For instance
$$
\eqalign{
\langle X\rangle&=\frac{1}{1+\hat f_1},\cr
\langle X^2\rangle&=\frac{1-\hat f_1}{(1+\hat f_1)(1-\hat f_2)},\cr
\langle X^3\rangle&=\frac{1-2\hat f_1+2\hat f_2-\hat f_1\hat f_2}
{(1+\hat f_1)(1-\hat f_2)(1+\hat f_3)},\cr
\cdots&
}
\eqno({\rm B.}1)
$$

\noindent (ii)
From (3.13 b), and using eq. (3.10), one has
$$
\langle\xi^k\rangle=\langle\e^{-\lambda\, k}\rangle
\langle X^k\rangle
=\frac{1-\hat f_k}{k\langle l\rangle}\langle X^k\rangle
.\eqno({\rm B.}2)
$$

\noindent (iii)
By symmetry, only even moments of $\m$ are non zero.
From (3.13 a) they are related to those of $\xi$ by a binomial expansion.

Finally one gets
$$
\eqalign{
\langle\m^2\rangle
&=1-\frac{2}{\langle l\rangle}\frac{1-\hat f_1}{1+\hat f_1},\cr
\langle\m^4\rangle
&=1-\frac{8}{3\langle l\rangle}
\frac{(1-2\hat f_1+2\hat f_2-\hat f_1\hat f_2)(1-\hat f_3)}
{(1+\hat f_1)(1-\hat f_2)(1+\hat f_3)}.\cr
\cdots&}
\eqno({\rm B.}3)
$$

Replacing $\hat f(s)$ by its expression in terms of $\hat A(s)$ (see eq. (3.6)),
permits to recast (B.3) into
$$
\eqalign{
\langle\m^2\rangle
&=\hat A_1\cr
\langle\m^4\rangle
&=1-
\frac{(1-3\hat A_1+4 \hat A_2)(1-3 \hat A_3)}
{1-2 \hat A_2}.\cr
}
\eqno({\rm B.}4)
$$
The first line of eq. (B.4) may be understood as follows.
In the long time regime, using the logarithmic time $\tau=\ln t$, one has
$$
\langle M^2(\tau)\rangle=2 \e^{-2\tau}\int_0^\tau \d \tau_2\e^{\tau_2}
\int_0^{\tau_2}\d \tau_1 \e^{\tau_1} A(\tau_2-\tau_1)
,\eqno({\rm B.}5)
$$
where $A(\tau)$ is the autocorrelation function (3.2).
Laplace transforming both sides of (B.5) gives
$$
\int_0^\infty \d \tau\, \e^{-s\tau}\langle M^2(\tau)\rangle
=\frac{2 \hat A(s+1)}{s(s+2)}
,\eqno({\rm B.}6)
$$
hence, when $\tau\to\infty$,
$$
\langle M^2\rangle=\hat A(1)
,\eqno({\rm B.}7)
$$
since the rightmost pole of the right hand side of (B.6) is at $s=0$.
The result (B.7) is generic for coarsening systems, whenever the autocorrelation
function is scaling in the two time variables. 
In particular it holds for the Ising chain studied in section 4.

\smallskip
We now show that the solution of eqs. (3.13 a, b, c) for $f(l)=\theta\e^{-\theta
l}$ is the beta law (\eqsing). 
Setting $Z=\e^{-l}$ in (3.13 c), the integral equation for the
invariant distribution $f_X$ reads
$$
f_X(x)=\int_0^1\d z\int_0^1\d y f_Z(z)\,f_X(y)\,\delta(x-1+zy)
,\eqno({\rm B.}8)
$$
where $f_Z$, the probability density function of the variable $Z$, is known from that
of the interval length $l$, $f(l)$.
For $f(l)=\theta\e^{-\theta l}$ one has
$f_Z(z)=\theta\,z^{\theta-1}$, which cast into eq. ({\rm B.}8) leads to the solution
$$
f_X(x)=B^{-1}(\theta+1,\theta)\, x^\theta\,(1-x)^{\theta-1}
,\eqno({\rm B.}9)
$$
where $B(\theta+1,\theta)$ is the beta function.
Then computing
$$
\langle\xi^s\rangle=\langle\e^{-\lambda\, s}\rangle
\langle X^s\rangle
,\eqno({\rm B.}10)
$$
with $\langle\e^{-\lambda\, s}\rangle=\hat q(s)=\theta/(s+\theta)$, and
$\langle X^s\rangle=B^{-1}(\theta+1,\theta)\,B(\theta+s+1,\theta)$ leads to 
$\langle \xi^s\rangle=B^{-1}(\theta,\theta+1)\,B(\theta+s,\theta+1)$ hence to the law
$$
f_\xi(x)=B^{-1}(\theta,\theta+1)\,x^{\theta-1}\,(1-x)^{\theta}
,\eqno({\rm B.}11)
$$ 
for the random variable $\xi$.
Finally, for this choice of $f(l)$, the solution of (3.13 a, b, c) is a beta law on
$(-1,1)$ 
$$
\f(x)=B^{-1}(1/2,\theta)\,(1-x^2)^{\theta-1}
,\eqno({\rm B.}12)
$$
which is eq. (\eqsing).


\vfill\eject
{\parindent 0em
{\bf References}
\vskip 12pt plus 2pt

[1] M. Marcos-Martin, D. Beysens, J.P. Bouchaud, C. Godr\`eche, and I. Yekutieli,
Physica A {\bf 214}, 396 (1995). 

[2] B. Derrida, A.J. Bray, and C. Godr\`eche, J. Phys. A {\bf 27}, L357 (1994).

[3] A.J. Bray, B. Derrida and C. Godr\`eche, Europhys. Lett. {\bf 27}, 175 (1994).

[4] E. Ben-Naim, P.L. Krapivsky, and S. Redner, Phys. Rev. E {\bf 50}, 2474 (1994).

[5] D. Stauffer, J. Phys. A {\bf 27}, 5029 (1994).

[6] J. Cardy, J. Phys. A {\bf 28}, L19 (1995). 

[7] B. Derrida, J. Phys. A {\bf 28}, 1481 (1995).

[8] B. Derrida, V. Hakim, and V. Pasquier, Phys. Rev. Lett. {\bf 75}, 751 (1995); 
J. Stat. Phys. {\bf 85}, 763 (1996).

[9] B. Derrida, P.M.C. de Oliveira, and D. Stauffer, Physica A {\bf 224}, 604 (1996).

[10] B. Derrida and V. Hakim, J. Phys. A {\bf 29}, L589 (1996).

[11] S.N. Majumdar, A.J. Bray, S.J. Cornell, and C. Sire, Phys. Rev. Lett. {\bf 77},
3704 (1996). 

[12] B. Derrida, V. Hakim and R. Zeitak, Phys. Rev. Lett. {\bf 77},
2871 (1996).

[13] S.N. Majumdar, C. Sire, A.J. Bray, and S.J. Cornell, Phys. Rev. Lett. {\bf 77},
2867 (1996).

[14] S.N. Majumdar and C. Sire, Phys. Rev. Lett. {\bf 77}, 1420 (1996). 

[15] C. Monthus, Phys. Rev. E {\bf 54}, 919 (1996). 

[16] E. Ben-Naim, L. Frachebourg, and P.L. Krapivsky, Phys. Rev. E {\bf 53}, 3078
(1996).

[17] B. Yurke, A.N. Pargellis, S.N. Majumdar and C. Sire, Phys. Rev. E {\bf 56}, R40
(1997).

[18] W.Y. Tam, R. Zeitak, K.Y. Szeto, and J. Stavans, Phys. Rev. Lett. {\bf 78}, 1588
(1997).

[19] B. Derrida, Phys. Rev. E {\bf 55}, 3705 (1997). 

[20] D. Stauffer, Int. J. Mod. Phys. C {\bf 8}, 361 (1997). 

[21] L. Frachebourg, P.L. Krapivsky, and S. Redner, Phys. Rev. E {\bf 55}, 6684
(1997).

[22] K. Oerding, S.J. Cornell and A.J. Bray, Phys. Rev. E {\bf 56}, R25 (1997). 

[23] P.L. Krapivsky and E. Ben-Naim, Phys. Rev. E {\bf 56}, 3788 (1997).

[24] J. Krug, H. Kallabis, S.N. Majumdar, S.J. Cornell, A.J. Bray, and C. Sire,
Phys. Rev. E {\bf 56}, 2702 (1997).

[25] N. Menyh\'ard and G. \'Odor, J. Phys. A {\bf 30}, 8515 (1997).

[26] B.P. Lee and A.D. Rutenberg, Phys. Rev. Let. {\bf 79}, 4842 (1997). 

[27] S. Cueille and C. Sire, J. Phys. A {\bf 30}, L791 (1997). 

[28] S.N. Majumdar and S.J. Cornell, {\it cond-mat} 9707344. 

[\refbgl] M. Bauer, C. Godr\`eche, J.M. Luck, unpublished.

[\refdk] D.A. Darling and M. Kac, Trans. Amer. Math. Soc. {\bf 84}, 444 (1957).

[\refgrif] D. Griffeath, {\it Additive and Cancellative Interacting Particle
Systems}, Springer Lecture Notes in Math. {\bf 724}, Springer (1979).

[\refcga] J.T. Cox and D. Griffeath, The Annals of Probability {\bf 11}, 876
(1983). 

[\refcgb] J.T. Cox and D. Griffeath, Z. Wahrsch. verw. Gebiete {\bf 66} 543 (1984).

[\refcgc] J.T. Cox and D. Griffeath, Contemporary Mathematics {\bf 41}, 55 (1985).

[\refbcg] M. Bramson, J.T. Cox and D. Griffeath, Probab. Th. Rel. Fields {\bf 77},
613 (1988).

[\refcox] J.T. Cox, Ann. Probab. {\bf 16}, 1559 (1988).

[\refkesta] H. Kesten, Acta Math. {\bf 131}, 208 (1973).

[\refkestb] H. Kesten, M.V. Kozlov and F. Spitzer, Compos. Math. {\bf 30}, 145
(1975).

[\refkestc] W. Vervaat, Adv. Appl. Prob. {\bf 11}, 50 (1979).

[\refkestd] C. de Calan, D. Petritis, J.M. Luck and Th.M. Nieuwenhuizen, J. Phys.
A {\bf 18}, 501 (1985).

[\refglau] R.J. Glauber, J. Math. Phys. {\bf 4}, 294 (1963).

[\reflig] T.M. Liggett, {\it Interacting Particle Systems}, Springer (1985).

[\refhg] M. Howard and C. Godr\`eche, {\it cond-mat} 9711148.

[\refdg] I. Dornic and C. Godr\`eche, in preparation.

[45] W. Feller, {\it Probability theory and its applications}, Wiley (1968).
}

\vfill\eject

\magnification\magstep 1
\def\init{\tabskip 0pt}
\def\crr{\cr\noalign{\hrule}}
\medskip
$$
\vbox{\init\halign to 15truecm
{\strut#&\vrule#\tabskip=1em plus 2em&
\hfil$#$\hfil&\vrule#&
\hfil$#$\hfil&\vrule#&
\hfil$#$\hfil&\vrule#&
\hfil$#$\hfil&\vrule#&
\hfil$#$\hfil&\vrule#\tabskip 0pt\crr
&&\ &&\ &&\ &&\ &&\ &\cr
&&\ &&a(t) &&b(t) &&I(x\to1) &&J(x\to1) &\cr
&&\ &&\ &&\ &&\ &&\ &\crr
&&\ &&\ &&\ &&\ &&\ &\cr
&&\hbox{Random walk}\hfill &&t &&t &&\ln 2 &&\ln 2 &\cr
&&\ &&\ &&\ &&\ &&\ &\crr
&&\ &&\ &&\ &&\ &&\ &\cr
&&\left\{\matrix{\hbox{Diffusion equation}\cr \hbox{1D Ising}\hfill}\right.
&&{\cal O}(1) &&\ln t &&{\rm \infty} && {\rm \theta} &\cr &&\ &&\
&&\ &&\ &&\ &\crr
&&\ &&\ &&\ &&\ &&\ &\cr
&&\hbox{2D Voter}\hfill &&\ln t &&(\ln t)^2 && {\rm \infty} &&{\rm const.} &\cr
&&\ &&\ &&\ &&\ &&\ &\crr
}}
$$

Table 1: Summary of results for the functions $a(t)$, $b(t)$, $I(x)$, $J(x)$
(see eqs. (2.4, 2.5)).

\magnification\magstep 1
\def\init{\tabskip 0pt}
\def\crr{\cr\noalign{\hrule}}
\medskip
$$
\vbox{\init\halign to 15truecm
{\strut#&\vrule#\tabskip=1em plus 2em&
\hfil$#$\hfil&\vrule#&
\hfil$#$\hfil&\vrule#&
\hfil$#$\hfil&\vrule#&
\hfil$#$\hfil&\vrule#&
\hfil$#$\hfil&\vrule#&
\hfil$#$\hfil&\vrule#&
\hfil$#$\hfil&\vrule#\tabskip 0pt\crr
&&\ &&\ &&\ &&\ &&\ &&\ &&\ & \cr
&&\ &&\langle M^2\rangle &&\mu_2 &&\langle M^4\rangle &&\mu_4 && \langle
M^6\rangle && \mu_6 &\cr 
&&\ &&\ &&\ &&\ &&\ &&\ &&\ &\crr
&&\ &&\ &&\ &&\ &&\ &&\ &&\ &\cr
&&\hbox{1D}\hfill &&.7996 &&.8060 &&.7383 &&.7462 &&.7035 &&.7119 &\cr
&&\ &&\ &&\ &&\ &&\ &&\ &&\ &\crr
&&\ &&\ &&\ &&\ &&\ &&\ &&\ &\cr
&&\hbox{2D} &&.7268 && .7286 && .6459 && .6482 && .6008 && .6032 &\cr
&&\ &&\ &&\ &&\ &&\ &&\ &&\ &\crr
}}
$$

Table 2: Values of the moments $\langle M^k\rangle$ for the diffusion equation,
computed in the independent interval approximation, compared to the moments of
the beta law (3.18).
\bye

\bye